\documentclass[aps,tighten,preprint]{revtex4}
\usepackage{graphicx}
\input epsf

\newcommand{\vk}{\vec{k}}

\newcommand{\mN}{m_{_N}}
\newcommand{\meta}{m_{_\eta}}
\newcommand{\vkapa}{\vec{\kappa}}

\newcommand{\sigf}{\Sigma^{free}}
\newcommand{\sigm}{\Sigma^{med} }
\newcommand{\siga}{\Sigma^{abs} }

\begin{document}
\title{Interference and nuclear medium effects on\\
the eta-mesic nuclear spectrum}
\author{Q. Haider\footnote{e-mail address: haider@fordham.edu}\\}
\address{Physics Department, Fordham University\\ Bronx, N.Y. 10458, USA}
\author{Lon-chang (L.C.) Liu\footnote{e-mail address: liu@lanl.gov\\}}
\address{Theoretical Division, Group T-2, Mail Stop B214\\
Los Alamos National Laboratory, Los Alamos, N.M. 87545, USA}

\begin{abstract}

The missing-mass spectrum obtained in a recoil-free transfer reaction
$p(^{27}$Al,${^3}$He)$\pi^{-}p'$X is analyzed. We find that the observed peak
structure arises from the coherent contributions from two reaction processes in the
energy region corresponding to a bound eta ($\eta$) meson.
In one of the processes the intermediate $\eta$ is captured by the nucleus
to form the $\eta$-mesic nucleus $^{25}$Mg$_{_{\eta}}$. In the other process,
the $\eta$ does not form $\eta$-nucleus bound state. The interference
between these two processes has caused the peak of
the spectrum to appear at an $\eta$ binding energy stronger than the actual one.
Our analysis also indicates that the data are consistent with an attractive
N$^{*}$(1535)-nucleus interaction at energies below the $\eta$N threshold.\\

\end{abstract}

\date{September 30, 2010} 
\maketitle

\section{Introduction}{\label {sec.1} }
The possibility of an eta ($\eta$) meson being bound in a nucleus by strong-interaction
force, leading to the formation of an exotic nucleus called  eta-mesic nucleus,
was first proposed in 1986 \cite{hai}. The nuclear binding of an $\eta$ meson is caused
by an attractive eta-nucleon ($\eta$N) interaction in the threshold region. This attractive
$\eta$N interaction is due to the N$^*$(1535) resonance being situated not too far above
the threshold of the $\eta$N channel (1488~MeV). It was also shown in Ref.\cite{hai} that
a bound state is possible if there is a sufficient number of nucleons (10 or more) in a nucleus.

The existence of $\eta$-mesic nuclei will certainly create new premises
for studying $\eta$ meson and baryon resonances inside nuclei. Extensive
interest in exploring this new venue is evidenced by the large amount of theoretical
work that has been done during the last two decades~\cite{kohn}-\cite{wilk}.
Irrespective  of the models and formalisms used, from a theoretical point of view
there is unanimity about the existence of $\eta$-mesic nucleus.

Recently, the COSY-GEM collaboration~\cite{Cosy} searched for $\eta$-mesic
nucleus by means of a recoil-free transfer reaction
$p(^{27}\mbox{Al},^{3}\mbox{He})\pi^{-}p'\mbox{X}$. The kinematics was so chosen that
the $\eta$ produced in the intermediate state is nearly at rest, favoring its
capture by the residual nucleus $^{25}$Mg. Because of energy conservation, the bound $\eta$ cannot
reappear as an observable particle in the  decay products of the mesic
nucleus $^{25}$Mg$_{_\eta}$. Instead, it
interacts, for example, with a target neutron resulting in the emission of
a nearly ``back-to-back" $\pi^{-}p$ pair in the laboratory.
We denote this multi-step reaction as Process M (M for mesic-nucleus):

 $p + \mbox{$^{27}$Al} \rightarrow  \underbrace{ \eta + \mbox{$^{25}$Mg}} + \mbox{$^{3}$He}$

  \hspace{1.125in}$\downarrow$

  \hspace{1.125in}$^{25}$Mg$_{_{\eta}}$

  \hspace{1.125in}$\downarrow$

  \hspace{.85in}$\overbrace{ \eta + \mbox{$^{25}$Mg}} \rightarrow (\pi^-+p) + $X\ \ .

In order to reduce a large number of background events arising from particles
being emitted during nuclear cascade process, the COSY-GEM collaboration
implemented a triple-coincidence measurement among $^3$He and the ``back-to-back"
$\pi^{-}p$ pair having the kinetic energy spectra of the pion and proton peaked,
respectively, at about 100 and 320~MeV~\cite{beti}. Because of this background reduction,
a peak in the missing-mass spectrum in the energy region corresponding to
bound $\eta$ has been made evident. Upon fitting the spectrum with the sum of
a background term and a Gaussian ($|f_{b}|^2 + |f_{g}|^2$), COSY-GEM collaboration determined
that the peak has its centroid situated at binding energy $(-13.13\pm 1.64)$~MeV with
a FWHM of $(10.22\pm 2.98)$~MeV [or a half-width $\Gamma/2\simeq (5.1 \pm 1.5)$ MeV].

By performing a bound-state calculation based on scattering length and using
on-shell kinematics, we find
that the above binding energy and half-width correspond to an effective $s$-wave
$\eta$N scattering length $a_{0}\simeq (0.292+0.077i)$~fm. An exceptional feature
of this scattering length is its imaginary-to-real part ratio
$R\equiv {\cal I}m(a_0)/{\cal R}e(a_0)$ = 0.26 only, while most of the published theoretical
models give scattering lengths (see table~\ref{tabI} of Ref.\cite{hai2}) having $R\gg 0.35$. In other
words, the value of $R$ given by the theories is higher than the fitted value by at least 35\%.
The need to understand the huge difference between theory and experiment has motivated the present study.

More specifically, we will reanalyze the experiment and infer from our analysis the nature
of the observed peak structure and the qualitative feature of the N$^{*}$(1535)-nucleus
interaction. In section~II we outline the mesic-nucleus theory to be used in the analysis.
Detailed analysis is given in section~III, and our findings are summarized in section~IV.

\section{Outline of the mesic-nucleus theory}{\label{sec.1.1}}
The eigenvalue equation of the bound-state of an $\eta$ meson in a nucleus is
$(H_0+V)|\psi\rangle={\cal E}|\psi\rangle$. The $\eta$-nucleus
potential $V$ is complex because the $\eta\rightarrow \pi$ channels are open.
Hence, the eigenenergy ${\cal E}$ is also complex and can be written as
${\cal E}=E_{bd}- i\Gamma/2$, where $E_{bd} \;(<0)$ and $\Gamma$ are
the binding energy and width of the bound state, respectively. The momentum-space
matrix elements of the leading-order potential are given by \cite{hai2}-\cite{hai3}
\begin{equation}
\langle \vk'\mid V \mid \vk\rangle =
\langle \vkapa'|t_{_{\eta N}}(W)
| \vkapa \rangle F(\vk'-\vk),
\label{1.1}
\end{equation}
where $\vk$ and $\vk'$ denote the initial and final $\eta$ momenta in the
$\eta$-nucleus c.m. frame and $F(\vk'-\vk)$ is the nuclear form factor.
The $t_{\eta N}$ is the operator for the scattering of $\eta$ from the nucleon.
The variables $\vkapa$, $\vkapa'$, and $W$ are, respectively,
the initial and final $\eta$N relative momenta, and the total
energy of the $\eta$N system in its c.m. frame.

Without loss of generality, we use the coupled-channel isobar (CCI) model of Bhalerao and Liu~\cite{bha}
to calculate the potential given by Eq.(\ref{1.1}). The reason for this is two-fold. First,
we have at our disposal the detailed energy dependence of the model which
reproduces remarkably the observable ($\pi$N S$_{11}$ phase shifts) in
the entire energy region where the nuclear binding of an $\eta$ could take place. Second, it was this
model that was used to predict the existence of $\eta$-nucleus bound states. Furthermore,
as has been noted in the previous section, the discrepancy between the theoretical and
experimental value of $R$ exists for all published models, including the CCI model.
Hence, we believe the general features of our findings are not limited to the model used.

In the CCI model of Ref.\cite{bha},
\begin{equation}
\langle \vkapa'|t_{\eta N}(W)|\vkapa\rangle
=K\sum_{\ell} v_{\ell}(\vkapa',\Lambda_{\ell}){\cal A}_{\ell}(W)
v_{\ell}(\vkapa,\Lambda_{\ell}),
\label{1.1b}
\end{equation}
where $K$ is a kinematic factor and $\ell=0,1,2$ are, respectively, the
$s$-, $p$-, and $d$-wave $\eta$N interactions. The $v_{\ell}$ are off-shell
form factors of range $\Lambda_{\ell}$, and ${\cal A}_{\ell}$ are the
energy-dependent amplitudes. For bound-state problems, $p$- and $d$-wave
interactions have negligible contributions. Consequently, we will only consider
the $s$-wave $\eta$N interaction and omit the subscript $\ell$.
The amplitude ${\cal A}$ is given by
\begin{equation}
{\cal A}(W) = \frac{g^2}{2W D(W)}.
\label{1.5}
\end{equation}
Conversely, if ${\tilde {\cal A}}$ is the amplitude that gives the measured
$E_{bd}$ and $\Gamma/2$, then Eq.(\ref{1.5}) can be used to derive the energy
dependence of the denominator, namely,
\begin{equation}
D(W) = \frac{g{^2}}{2W{\tilde{\cal A}}}.
\label{1.5a}
\end{equation}
In the above equations
$g$ is the $\eta$NN$^{*}$ coupling constant. Here, N${^*}$ is the
$s$-wave isobar N$^*(1535)$ which has a mass between
1525 and 1545~MeV and a Breit-Wigner width from
125 to 175~MeV~\cite{Hand}. For mesic-nucleus calculations,
\begin{equation}
W = \meta + \mN + \langle B_{{_N}}\rangle ,
\label{1.5b}
\end{equation}
where $\langle B_{{_N}}\rangle <0$ is the average binding energy of the nucleon.
In Eq.(\ref{1.5})
\begin{equation}
 D(W) = W -  M_{N^*}(W),
\label{1.6}
\end{equation}
and
\begin{equation}
 M_{N^{*}}(W) =
 M^{0} + r(W) + {\cal R}e[\sigm(W)]
 + i\ {\cal I}m[\sigm(W)],
\label{1.7}
\end{equation}
with $\sigm=\sigf+\siga$. The $\sigf$ is the N$^{*}$ self-energy arising from
its decays to the $\eta$N, $\pi$N, and $\pi\pi$N channels in free space.
If $\gamma^{free}(W)$ denotes its total free-space decay width, then
$\frac{1}{2}\gamma^{free}(W) =\  -{\cal I}m[\sigf(W)]$.
The bare mass $M^0$, coupling constant $g$, range parameter $\Lambda$
needed for the calculation of $\sigf$ were
determined from fitting the experimental $\pi$N S$_{11}$ phase shifts~\cite{bha}.
The $\Sigma^{abs}$ is the N$^*$ self-energy arising from true absorption
(or annihilation) by nucleons of the pions coming from N$^*\rightarrow \pi$N,
and $\pi\pi$N decays. To our knowledge, microscopic absorption model that can fit
systematically all experimental data is still not available. In the literature,
${\cal I}m[\siga]$ has been estimated in the framework of  local-density approximation,
and ${\cal R}e[\siga]$ is treated as a parameter~\cite{cha}. It was found that
total $-{\cal I}m[\siga] \simeq 35$~MeV at $W=1535$~MeV
and at nuclear density $\rho=\rho_{_{0}}=0.17$~fm$^{-3}$.
We extend this result to the subthreshold region by using
$-{\cal I}m[\Sigma^{abs}] = 35(q/^{\sim}\!\!\! q )^3(\rho/\rho_{_{0}})^{2}$~MeV.
Here, $^{\sim}\!\!\! q$ is the $\pi$N relative momentum at
$W=1535$~MeV and $q$ the corresponding one at $W < (\mN+\meta )$.
The exponent of $q$ is based on the local-density result~\cite{cha} on the momentum
dependence of $\siga$ while the exponent of $\rho$ is based on the fact that in a nucleus,
pion absorption involves at least two nucleons.
In principle, any subthreshold N${^*}$-nucleus interactions
can contribute to the real part of the N${^*}$ self-energy.
We denote these contributions collectively as $r(W)$
in Eq.(\ref{1.7}).

\begin{figure}
\includegraphics[angle=0,width=0.5\columnwidth]{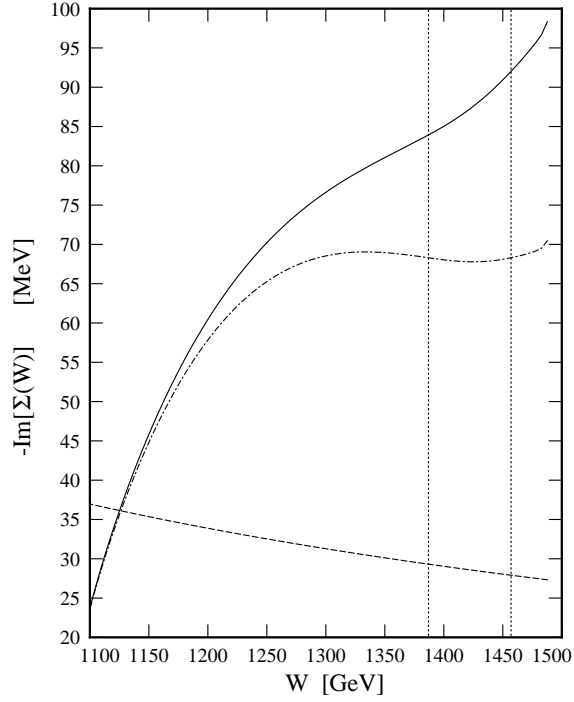}\\
\caption{The energy dependence of $-{\cal I}m[\sigf]$ (dash-dotted curve)
and $- \left ({\cal I}m[\sigf]\right.$ $+ \left. {\cal I}m[\siga]_{\rho=\rho_{_{0}}}\right )$ (solid curve).
The left and right vertical dotted lines indicate, respectively, the $W$ corresponding to
$\langle B_{_{N}}\rangle = -100$ and $-30$~MeV. The dashed curve gives the energy dependence
of $-{\cal I}m[\sigm]$ in case the experimental spectrum [17] is due solely
to mesic-nucleus formation (see the text in Section~\ref{sec.2}).}
\label{Fig1}
\end{figure}

We emphasize that ${\cal I}m[\siga]$ has the same sign
as ${\cal I}m[\sigf]$, as required by the unitarity of an optical potential.
Hence, $\ | {\cal I}m[\sigm]\ | \geq \ | {\cal I}m[\sigf] \ |$.
The dependence of $\sigf$ and the maximal $\sigm$ on $W$ are
shown, respectively, as the dash-dotted and solid curves in fig.{\ref{Fig1}}.
At any given $W$, a physically meaningful ${\cal A}(W)$ must always yield an
${\cal I}m[\Sigma^{med}]$ situated in the ``physical zone" bordered by
these two curves, which we term the unitarity requirement.
In fig.{\ref{Fig1}} the left and right vertical lines show the positions of
$W$ corresponding, respectively, to nucleon binding energies
$\langle B_{_{N}}\rangle= -100$ and $-30$~MeV. As one can see,
between these two $\langle B_{_{N}}\rangle$'s the lower boundary
of ${\cal I}m[\sigm]$ is nearly constant.

The quantity $r+{\cal R}e[\siga]$ in Eq.(~\ref{1.7}) represents the real part of the N$^*$-nucleus
interaction which we denote as $ V_{N^{*}}$. Equation~(\ref{1.7}) can then be written as
\begin{equation}
M_{N^{*}}(W) =
M^{0} + V_{N^{*}}(W) + {\cal R}e[\sigf(W)]
+ i\ {\cal I}m[\sigm(W)] \ .
\label{1.7b}
\end{equation}
Using Eq.(\ref{1.7b}) in Eq.(\ref{1.6}), we obtain
\begin{equation}
D(W)= W - \left(\  M^0 + V_{N^{*}}(W) + {\cal R}e[\sigf(W)]  +\
i\ {\cal I}m[\Sigma^{med}(W)]\ \right) \ .
\label{1.10}
\end{equation}
Upon equating the real and imaginary parts of Eq.(\ref{1.10}), one obtains
\begin{equation}
V_{N^{*}}(W) = W - {\cal M}(W) -  {\cal R}e[D(W)],
\label{1.16}
\end{equation}
and
\begin{equation}
{\cal I}m[\sigm(W)] = -{\cal I}m[D(W)].
\label{1.17}
\end{equation}
In Eq.(\ref{1.16}), ${\cal M}(W)\equiv M^0 + {\cal R}e[\sigf(W)]$ and $W$ is given
by Eq.(\ref{1.5b}). We recall that
$\langle B_{_{N}}\rangle$ is the average nucleon binding energy. Because the nucleons
are bound, $\langle B_{_{N}} \rangle < 0.$ On the other hand, $V_{N^{*}}$ can be
either negative or positive, depending on whether the interaction is attractive or
repulsive.

\section { Analysis and Discussion}{\label {sec.2}}
If the centroid of the experimental peak at $-13$~MeV and half-width
of 5~MeV are due solely to the formation of the $\eta$-mesic nucleus
$^{25}$Mg$_{_\eta}$ via Process M, then an amplitude
$\tilde{\cal A}=-(0.0521+0.0099i)$~fm${^2}$ is required to reproduce the
above data. Upon using Eqs.(\ref{1.5a}) and (\ref{1.17}) to solve for
${\cal I}m[\sigm(W)]$, we obtain the dashed curve in fig.{\ref{Fig1}}.
As one can see, this curve intersects the physical zone
at $W \simeq 1125$~MeV, which, by Eq.(\ref{1.5b}), corresponds to
a $\langle B_{_{N}}\rangle \simeq -360$~MeV. This is clearly an
unrealistic value which we regard as a strong indication that Process M alone is insufficient
in describing the observed spectrum.

Indeed, the $\eta$ produced in the intermediate state can also
be scattered by the residual nucleus and emerge as a pion,
without being first captured by the nucleus.
We denote this multi-step reaction as Process S (S for scattering):

 $p + \mbox{$^{27}$Al} \rightarrow  \underbrace{ \eta + \mbox{$^{25}$Mg}} + \mbox{$^{3}$He}$

 \hspace{1.125in}$\downarrow$

 \hspace{.85in}$\overbrace{ \eta + \mbox{$^{25}$Mg}} \rightarrow (\pi^-+p) + $X\ \ .

\begin{figure}
\includegraphics[angle=0,width=0.5\columnwidth]{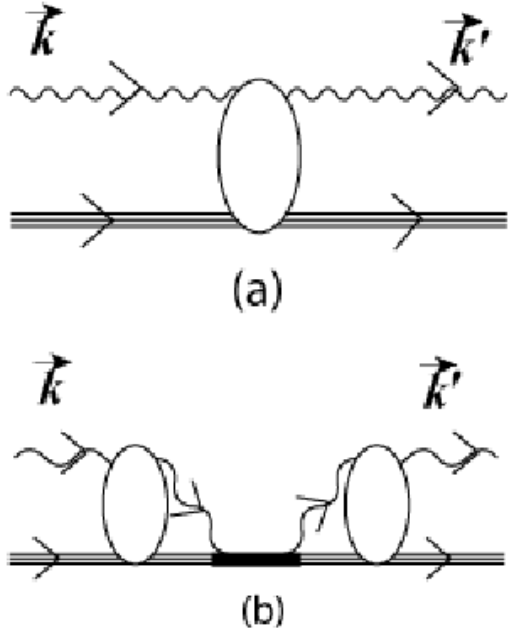}\\
\caption{(a) Reaction diagram of $f_{_S}$. (b) Reaction diagram of $f_{_M}$.
The wavy and multiple lines represent, respectively, the $\eta$ and $^{25}$Mg.
The open oval denotes the $\eta$-nucleus interaction $V$. The filled line in (b)
denotes the mesic nucleus. }
\label{Fig2}
\end{figure}

The essential portion of the reaction dynamics that differentiates the S and M processes,
as indicated by upper and lower braces in the corresponding reaction equations,
are illustrated in fig.{\ref{Fig2}}. We emphasize that because these two reaction paths lead to the same
measured final state, they cannot be distinguished by the experiment. Consequently, in theoretical analysis
one must take coherent summation of the two amplitudes to account for the quantum interference
between them. We, therefore, fit the experimental spectrum by using the sum of two amplitudes:
\begin{equation}
\alpha\mid f_{_S} + f_{_M} \mid^2 =
\alpha\left|<\vec{k}'|V(E)|\vec{k}> +
\frac{<\vec{k}'|V(E)|\psi> <\Psi|V(E)|\vec{k}>}{E-(E_{bd}-i\Gamma /2)}\right| ^{2},
\label{1.21}
\end{equation}
where $V$ is given by Eq.(\ref{1.1}), $E\equiv W-m_{_{\eta}}-m_{_{N}}-\langle B_{N}\rangle$, $\psi$ is the wave function of
bound $\eta$, and $\Psi$ is its adjoint~\cite{rod}. We have noted that in the threshold
and subthreshold regions, $\eta$-nucleus interaction is isotropic and that the matrix
elements $<k'|V(E)|k>$ are nearly constant for $k$ and $k'$ between 0 and
100~MeV/$c$. Because of these aspects of the $\eta$-nucleus interaction and the experimental selection of
events corresponding to $\eta$ being produced nearly at rest, Eq.(\ref{1.21}) can be evaluated at
$|\vec{k}|=|\vec{k}'|\simeq 0$.

One should note that in Eq.(\ref{1.21}), there is only one parameter $\alpha$ and its role is to
just adjust the overall magnitude. We emphasize that we  used the same $V$ in calculating
$f_{_{S}}$ and $f_{_{M}}$, and that the values of $E_{bd}$ and $\Gamma /2$ obtained with this $V$
were kept fixed during the fit. Furthermore, we square the sum of the amplitudes, in marked
contrast to using the sum of the squared individual amplitudes. Hence, interference effects
between the amplitudes are present in our analysis while they are absent in the COSY-GEM fit.

Our main goal is to find answer to the following questions: (1)Would it be possible to explain
the spectrum by having $|E_{bd}|\ll$13~MeV in
$f_{_{M}}$? (2) Is $V_{N^{*}}$ attractive or repulsive?

\begin{figure}
\includegraphics[angle=0,width=0.5\columnwidth]{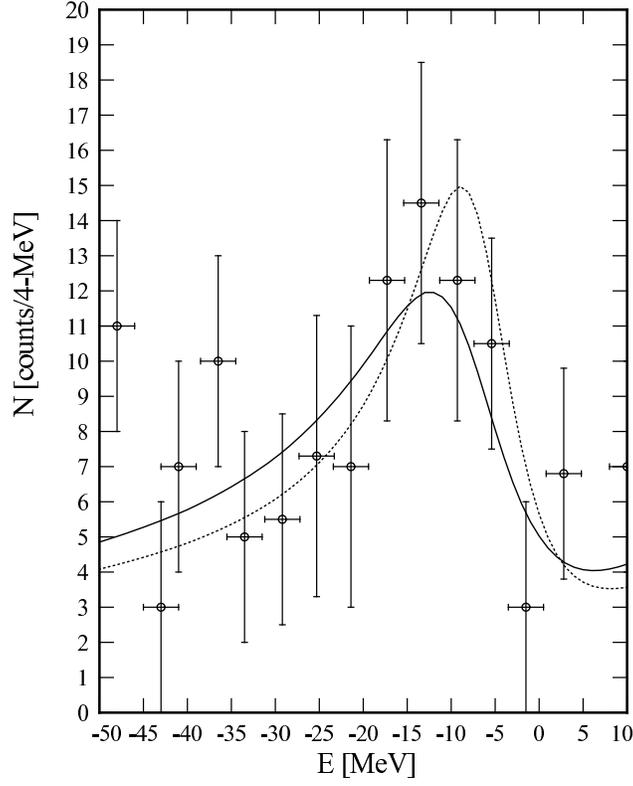}\\
\caption{Spectra obtained with (a) potential $V$ giving $E_{bd}-i\Gamma /2 = -(6.5 + 7.1i)$~MeV (dashed curve),
and (b) potential $V$ giving $E_{bd}-i\Gamma /2 =-(8.0 + 9.8i)$~MeV (solid curve). The data
are from Ref.\cite{Cosy}.} \label{Fig3}
\end{figure}

In our previous work~\cite{hai2}, we predicted the existence of $\eta$-mesic nuclei
with $<B_{_{N}}>=-30$~MeV, but without the inclusion of N$^{*}$-nucleus interaction.
In other words, we have set both $V_{N^{*}}$ and ${\cal I}m|\Sigma^{abs}|$ equal to zero.
The use of average nucleon binding energy $<B_{_{N}}>=-30$~MeV is based on the findings
from various studies of meson-nucleus scattering~\cite{liu,cott}. Applying the same
approach to $^{25}$Mg, we obtained ${\cal E}=E_{bd}-i\Gamma /2 = -(6.5+7.1i)$~MeV. Upon
using this ${\cal E}$ and the corresponding $V$ to calculate $f_{_{M}}$ and
$f_{_{S}}$, we obtained the result given in row (a) of table~\ref{tabI}
and the spectrum is displayed as the dotted curve in fig.{\ref{Fig3}}. As one can see, the peak position
of this curve appears at $-10.5$~MeV. This 4.0-MeV downward shift from $-6.5$~MeV indicates clearly the
importance of interference effect. With the use of $V$ obtained with $V_{N^{*}}=-24$~MeV and
$-{\cal I}m|\Sigma^{abs}|=0.65$~MeV, we obtained $E_{bd}=-8.0$~MeV and $\Gamma /2=9.8$~MeV.
The corresponding spectrum given by Eq.(\ref{1.21}) is shown as solid curve in fig.{\ref{Fig3}}.
The peak position of this curve is at $-12.5$~MeV, which agrees with the COSY-GEM result of
$(-13.13\pm 1.64)$~MeV. The quantitative aspect of this latter calculation is given in row
(b) of table~\ref{tabI}.

It is interesting to note from table~\ref{tabI} that fits (a) and (b) give same value to the
overall conversion parameter $\alpha$. As indicated in section~II, the $\eta$-nucleus
interaction strength is given by $D(W)$ and its values are shown in the 6th and 7th columns of
the table. The comparison of the 7th and 8th columns of the table shows that
${\cal I}m[D] (\equiv - {\cal I}m[\sigm]) \geq - {\cal I}m[\sigf]$. Hence, they satisfy the
unitarity requirement. We also did calculations using larger ${\cal I}m[D]$. However, they
led to larger calculated half-widths which worsened the fit. Consequently, we conclude
that the present data require ${\cal I}m[\sigm]$ being close to the lower boundary of the
physical zone of the self-energy (the dashed-dotted curve in fig.{\ref{Fig1}}). This in turn implies
that in the recoilless transfer reaction of Ref.\cite{Cosy},
${\cal I}m[\siga] \left ( ={\cal I}m[\sigm] - {\cal I}m[\sigf] \right )$ is very small,
and pion absorbtion takes place mainly in the
nuclear surface region where the condition $\rho /\rho_{0}\ll 1$ leads to a very small
${\cal I}m|\Sigma^{abs}|$.

\begin{table}
\begin{center}
\caption{Quantitative details ($\alpha$ is in counts/fm$^{2}$; all other quantities are in MeV).}
\label{tabI}

\medskip
\begin{tabular}
{|c|c|c|c|c|c|c|c|c|c|}\hline\hline
Fit & $\alpha$ & $E_{bd}$ & $\Gamma$/2 & $<B_{_{N}}>$ & ${\cal R}e[D]$ & ${\cal I}m[D]$ &
$-{\cal I}m[\sigf]$ & ${\cal M}$ & $V_{N^{*}}$ \\
\hline
(a)  & 4.2 & $-6.5$ & 7.1  & $-30$ & $-164$ & 68.3 & 68.3 & 1622 & 0 \\
(b)  & 4.2 & $-8.0$ & 9.6  & $-30$ & $-140$ & 69.0 & 68.3 & 1622 & $-24$ \\ \hline
\end{tabular}
\end{center}
\end{table}

\section{Conclusions}{\label {sec.3} }
Our analysis shows that two reaction processes are contributing to the
observed spectrum of the bound $\eta$ in $^{25}$Mg. The interference between these
processes causes the centroid of the observed spectrum to appear at an energy stronger
than the actual binding energy of the $\eta$ meson. The effective $\eta$N scattering
lengths that reproduced $E_{bd}$ and $\Gamma /2$ used in $f_{_M}$ associated with the
dashed and solid curves of fig.{\ref{Fig3}} are, respectively, $(0.226 +0.094i)$~fm and
$(0.250 +0.123i)$~fm. The corresponding imaginary-to-real part ratios are $R$=0.42 and 0.49,
consistent with theories (see Section~\ref{sec.1}). We, therefore, explained the apparent
discrepancy between theory and experiment.

The present analysis also indicates that the real part of the
interaction between N$^*$ and $^{25}$Mg is attractive at energies below
the $\eta$N threshold. This latter new nuclear information should be of value
to nuclear physics studies involving the baryon resonance N$^*$(1535) in
medium-mass nucleus, such as $^{25}$Mg.

We emphasize that the existence of S and M processes and the interference between them
are of a general nature. Consequently, our finding on the effects
arising from this aspect of reaction dynamics is model-independent.
On the other hand, while the specific value of  $V_{N^{*}}$
may be model-dependent, its sign (or the attractive nature of the interaction)
is model-independent. This is because the negative sign is required to provide
more binding which, when combined with interference effects, can lead to sufficient
downward shift of the peak in the binding-energy spectrum. We invite other
researchers having at their disposal the detailed off-shell properties of their
models to analyze the COSY-GEM data to further pin down the value of $V_{N^{*}}$.

\end{document}